\documentclass[review,12pt]{elsarticle}
\usepackage{lineno,hyperref}
\modulolinenumbers[5]

\journal{Journal of \LaTeX\ Templates}

\usepackage{amsmath}
\usepackage{graphicx}
\usepackage{mathtools}

\title{ High Frequency Conductivity of Hot Electrons in Carbon Nanotubes   }

\author[els]{M. Amekpewu\corref{cor1}}
\author[rvt]{S. Y. Mensah}
\author[els]{R. Musah}
\author[focal]{ N. G. Mensah}
\author[rvt]{\\ S. S. Abukari}
\author[rvt]{ K. A. Dompreh}
\address[els]{Department of Applied Physics, University for Development Studies,
Navorongo, Ghana}
\address[rvt]{Department of Physics, College of Agriculture and Natural Sciences, U.C.C, Ghana.}
\address[focal]{Department of Mathematics, College of Agriculture and Natural Sciences, U.C.C, Ghana}
\cortext[cor1]{Corresponding author\\ Email: mamek219@gmail.com}
\ead[url]{mamek219@gmail.com}

\date{}

\begin{document}
\begin{abstract}
High frequency conductivity of hot electrons in an undoped single walled achiral carbon nanotubes (CNTs) 
under the influence of ac-dc driven fields is considered. We investigated semiclassically by solving the 
Boltzmann's transport equation with and without the presence of the hot electrons source to derive the 
current densities. Plots of the normalized current density versus frequency of ac-field  reveal an 
increase in both the minimum and maximum peaks of normalized current density at lower frequencies as a 
result of  a strong enough injection of hot electrons . The applied ac-field plays twofold role of 
suppressing the space-charge instability in CNT and simultaneously pumping an energy for  lower frequency 
generation and amplification of THz radiations which have enormous promising applications in very different 
areas of science and technology.\\
Keywords: Carbon Nanotube, Hot electrons, conductivity, high frequency
\end{abstract}

\maketitle
\section*{Introduction}
Carbon nanotubes (CNTs)~\cite{1, 2, 3} have been the subject of intense research~\cite{4}-\cite{18}, since   
discovery in $1991$ by the Japanese scientist Sumio Iijima. Their unique structures, fascinating 
electronic, magnetic and transport properties have  sparked the interest and imagination of researchers 
worldwide~\cite{19}. These quasi-one-dimensional monomolecular nanostructural materials have a wide variety 
of possible applications~\cite{20}-\cite{22}.  
Research in hot electrons, like any field in semiconductor research, has received a great deal of 
attention since the arrival of the transistor in $1947$~\cite{23}. Recently, it has become possible to 
fabricate semiconductor devices with submicron dimensions. The miniaturization of devices has led 
to high field well outside the linear response region, where Ohm's law holds for any reasonable 
voltage signal~\cite{24}. The physical understanding of the microscopic processes which underlie the 
operations of such devices at high electric fields is provided by research into hot electron 
phenomena~\cite{25}. 
Whereas, there are several reports on hot electrons generation in CNTs~\cite{26, 27, 28, 29}, the reports on  
high frequency conductivity of hot electrons in CNTs are limited. Thus, in this paper, we present a 
theoretical framework investigations of high frequency conductivity of hot electrons in ($3,0$) 
zigzag ($zz$) CNT and ($3,3$) armchair ($ac$) 
The Boltzmann transport equation is solved in the framework of momentum-independent relaxation time 
using the semiclassical approach to obtain current density for each achiral CNTs. We probe the behavior 
of the electric current density of the CNTs as a function of the frequency  of ac field with and without 
the axial injection of the hot electrons.

\section*{Theory}
If a dc field $E_z$ is applied along a $z-$ axis of an undoped single-wall  carbon nanotube, electrons 
begin to move in accordance with the semiclassical Newton's law (neglecting scattering)~\cite{30}
\begin{equation}
\frac{dP_z}{dt} = eE_z
\end{equation}
where $P_z$ and $e$ are the axial component of the quasimomentum and the electronic charge of the 
propagating electrons respectively. For a CNT, If energy level spacing 
$\Delta \varepsilon$ ( $\Delta \varepsilon = \pi \hbar V_F /L $, $\hbar = h/2\pi$, 
$h$ is Planck constant, $V_F$ is Fermi velocity and $L$ 
is the length of the nanotube) is large enough and  the scattering rate $v$ is small such that                    
$\Delta \varepsilon >>aeE_z$ and $hv < aeE_z$ ($zz-CNT$), and $\Delta \varepsilon >>\frac{a}{\sqrt{3}} aE_z$
$h v < \frac{a}{\sqrt{3}}aE_z$, ($ac-CNT$), then the electrons oscillate inside the lower level with 
so-called Bloch frequency $\Omega$ given by~\cite{31}: 
\begin{equation}
\Omega_{zz} =\frac{aeE_z}{\hbar}
\end{equation}
\begin{equation}
\Omega_{ac} =\frac{aeE_z}{\sqrt{3}\hbar}
\end{equation}
for $zz-$CNT and $ac-$CNT respectively.  Here, $a$ is the lattice constant of the CNT. The investigation is 
done within the semiclassical approximation in which the motion of the $\pi$-electrons are considered as 
classical motion of free quasiparticles in the field of the crystalline lattice with dispersion law 
extracted from the quantum theory. Taking into account the hexagonal crystalline structure of a rolled 
graphene in a form of CNTs and using the tight binding approximation, the  energies for $zz-$CNT and $ac-$CNT 
are expressed as in equations ($4$) and ($5$), respectively~\cite{31}
\begin{eqnarray}
\varepsilon(s\Delta{p_{\phi}},p_z) &\equiv&  \varepsilon_s(p_z)=\nonumber\\
&\pm&\gamma_0 \sqrt{1 + 4cos(a{p_z})cos(\frac{a}{\sqrt{3}}s\Delta p_{\phi}) + 4cos^2(\frac{a}{\sqrt{3}}s\Delta p_{\phi})}
\end{eqnarray}

\begin{eqnarray}
\varepsilon(s\Delta{p_{\phi}},p_z) &\equiv&  \varepsilon_s(p_z)=\nonumber\\
&\pm&\gamma_0 \sqrt{1 + 4cos(as \Delta {p_{\phi}})cos(\frac{a}{\sqrt{3}} p_{z}) + 4cos^2(\frac{a}{\sqrt{3}} p_{z})}
\end{eqnarray}
where $\gamma_0\approx 3.0eV$ is the overlapping integral, $p_z$ is the axial component of quasimomemtum. 
$\Delta p_{\phi}$  is transverse quasimomentum level spacing and $s$ is an integer.  The expression for  
lattice constant a in equations ($4$) and ($5$) is given by    
\begin{equation}
a =\frac{3a_{c-c}}{2\hbar}
\end{equation}
where $a_{(c-c)}= 0.142nm$ is the C-C bond length. The $-$ and $+$ signs correspond to the valence and 
conduction bands respectively. Due to the transverse quantization of the quasimomentum  $P$, its transverse 
component $p_{\phi}$ can take $n$ discrete values, 
\begin{equation}
p_{\phi} =s\Delta p_{\phi} =\frac{\pi \sqrt{3}s}{an} \quad \quad (s = 1, ....,n)
\end{equation}
 Unlike transverse quasimomentum, $p_{\phi}$, the axial quasimomentum $\phi_z$ is assumed to vary continuously within 
the range $0 \leq {p_z}\leq 2\pi/a$, which corresponds to the model of infinitely long CNT ($L=\infty$). 
This model is applicable to the case under consideration because we are restricted to temperatures 
and/or voltages well above the level spacing~\cite{32}, i.e. $k_{\beta}T>\varepsilon_c,\Delta \varepsilon$,
where $k_{\beta}$ is Boltzmann constant, $T$ is the temperature, $\varepsilon_c$ is the charging energy. 
The energy expression in eq ($4$) and ($5$) can be expressed in the Fourier series as 
\begin{equation}
\varepsilon(p_z,s\Delta p_{\phi}) = \varepsilon(p_z) = \gamma_0\sum_{r\neq 0}{exp(iarp_z)}
\end{equation}
where $\varepsilon_{rs}$ is given as 
\begin{equation}
\varepsilon_{r,s} = \frac{a}{2\pi\gamma_0}\int_0^{\frac{2\pi}{a}}{\varepsilon_s(p_z)exp(-irap_z)}dp_z
\end{equation}
the quasiclassical velocity  of an electron moving along the CNTs axis is 
given by the expression $v_z (p_z, s\Delta p_{\phi}) =\partial\varepsilon_{rs}(p_z)/\partial p_z $. Substituting
eqn ($9$) and expressing further gives 
\begin{equation}
v_z(p_z,s\Delta p_z) = \gamma_0\sum_{r\neq 0}{\frac{\partial(\varepsilon_{rs}exp(iarp_z)}{\partial p_z}}
=\gamma_0\sum_{r\neq 0 }{iar\varepsilon_{rs}exp(iarp_z)}
\end{equation}
Considering the presence of hot electrons source, the motion of quasiparticles in an external axial 
electric field is described by the Boltzmann kinetic equation in the form as  shown  below~\cite{30, 31}
\begin{equation}
\frac{\partial f(p)}{\partial t} + v_z\frac{\partial f(p)}{\partial x} +    eE(t)\frac{\partial f(p)}{\partial p_z}  = -\frac{f(p) - f_0(p)}{\tau} + S(p)
\end{equation}
where $S(p)$ is the hot electron source function,  $f_0 (p)$ is equilibrium Fermi distribution 
function, $f(p,t)$ is the distribution function, $v_z$ is the quasiparticle group velocity along the $z-$axis of carbon nanotube  
and $\tau$ is the relaxation time. The relaxation term of equation ($11$)  describes the electron-phonon 
scattering, electron-electron collisions~\cite{31, 32} etc. Using the method originally developed in the 
theory of quantum semiconductor superlattices~\cite{31}, an exact solution of equation ($11$) can be 
constructed without assuming a weak electric field.  Expanding the distribution functions of interest 
in Fourier series as
\begin{equation}
f(p,t) = \Delta p_{\phi} \sum_{s=1}^n{\delta(p_{\phi} - s\Delta p_{\phi})}
\sum_{\tau \neq 0}{f_{rs}exp({iarp_z})\psi_v (t)}
\end{equation}
\begin{equation}
f_0(p) = \Delta p_{\phi} \sum_{s=1}^n{\delta(p_{\phi} - s\Delta p_{\phi})}
\sum_{\tau \neq 0}{f_{rs}exp({iarp_z})}
\end{equation}
for $zz$-CNTs
\begin{equation}
f(p,t) = \Delta p_{\phi} \sum_{s=1}^n{\delta(p_{\phi} - s\Delta p_{\phi})}
\sum_{\tau \neq 0}{f_{rs}exp({ibrp_z})\psi_v(t)}
\end{equation}
\begin{equation}
f_0(p) = \Delta p_{\phi} \sum_{s=1}^n{\delta(p_{\phi} - s\Delta p_{\phi})}
\sum_{\tau \neq 0}{f_{rs}exp({ibrp_z})}
\end{equation}
for $ac$-CNTs
where $b = a/\sqrt{3}$ or $a = b/\sqrt{3}$, $\delta(p_{\phi}-s\Delta p_{\phi})$ is the Dirac-delta 
function, $f_{rs}$ is the coefficients of the Fourier series and $\psi_v(t)$ is the factor by which 
the Fourier transform of the nonequilibruim distribution function differs from its equilibrium 
distribution counterpart. The expression for $f_{rs}$ can be expanded in the analogous form as 
\begin{equation}
f_{rs} =\frac{a}{2\pi}\int_{0}^{\frac{2\pi}{a}}{\frac{exp(-iarp_z)}{1+exp(\varepsilon_s(p_z))/k_{\beta}T)}}dp_z
\end{equation}
 The electron surface current density $j_z$ along the CNTs axis is also given by 
the expression
\begin{equation}
j_z =\frac{2e}{(2\pi\hbar)^2}\int\int{f(p,t)v_z(p)d^2p}
\end{equation}
the integration is carried over the fist Brillouin zone. 
For simplicity, we consider a hot electron source of the simplest form given by the expression,
\begin{equation}
S(p) = \frac{Qa}{\hbar}\delta(\phi-\phi^\prime) - \frac{aQ}{n_0}f_s(\phi)
\end{equation}
where $f_s (p)$ is the stationary (static and homogeneous) solution of equation ($19$), $Q$ is 
the injection rate of hot electron , $n_0$ is the equilibrium particle density, $\phi$ and $\phi^{\prime}$
 are the dimensionless momenta of  electrons and hot electrons respectively which are expressed as 
$\phi_{zz} = ap_z/\hbar$ and $\phi_{zz}^\prime = a p_z^\prime/\hbar$ for $zz$-CNTs and 
$\phi_{ac} = ap_z/{\sqrt{3}\hbar}$ and $\phi_{ac}^\prime = ap_z^\prime/\sqrt{3}\hbar$ for $ac$-CNTs,
We now find the high frequency conductivity of hot electrons in the nonequilibrium  state for $zz$-CNT by 
considering perturbations with frequency $\omega$ and wave-vector $k$ of the form 
\begin{equation}
E(t) = E_z + E_{\omega,k}exp(-i\omega t + ikx)
\end{equation}
\begin{equation}
f =f_s(\phi) + f_{\omega,k}exp(-i\omega t +ikx)
\end{equation}
Substituting equations ($19$)  and ($20$) into equation ($11$) and rearranging yields,           
\begin{equation}
\frac{\partial f_{\omega,k}}{\partial \phi} + i[\alpha + kv_z]f_{\omega,k} = -\frac{E_{\omega,k}}{E_z}
\frac{\partial f_s(\phi)}{\partial \phi}
\end{equation}
where $\alpha = -(\omega + iv_z)\ \Omega_{zz}$. Solving the homogeneous differential equation ($21$) 
and then introducing the Jacobi-Anger expansion and averaging the current over time, we obtain the current density for the $zz$-CNTs
in the presence of hot electrons ($j_{zHE}^{zz}$) as 
\begin{eqnarray}
j_{zHE}^{zz} = i\frac{4\sqrt{3}e^2 \gamma_0}{n\hbar^2} \sum_{l=1}{r} \sum_{s=1}{f_{rs}\varepsilon_{rs}}
\sum_{m,l=-\infty}{\frac{i^l mlj_m(\beta)j_{m-l}(\beta)I_{m-l}(\beta)\Omega_{zz}}{\omega + iv - m\Omega_{zz}}}\times\nonumber\\
\eta_{zz}\frac{n_0}{2\pi}\sum_{r}\frac{\Omega_{zz}exp(ir\phi)}{(ir\Omega_{zz} + v +\eta\Omega_{zz})}
( exp(-ir\phi^{\prime} - \frac{v}{(v + ir\Omega_{zz})}) + \frac{v}{(v + ir\Omega_{zz})})
\end{eqnarray}
where $\beta$ is the normalized amplitude of the ac-field, $j_m(\beta)$ is the bessel function order $m$ and 
$I_m(\beta)$ is the modified bessel function order $m$. In the absence of hot electrons, the nonequalibrium parameter for $zz$-CNT  $\eta_{zz} = 0$, hence 
the current density for $zz$-CNTs without hot electron source $j_(z )^{zz}$  could be obtained 
from equation ($22$) by setting $\eta_{zz} = 0$. Therefore, the current density of $zz$-CNTs in 
the absence of hot electrons $j_z^{zz}$ is given by 
\begin{equation}
j_{z}^{zz} = i\frac{4\sqrt{3}e^2 \gamma_0}{n\hbar^2} \sum_{l=1}{r} \sum_{s=1}{f_{rs}\varepsilon_{rs}}
\sum_{m,l=-\infty}{\frac{i^l mlj_m(\beta)j_{m-l}(\beta)I_{m-l}(\beta)\Omega_{zz}}{\omega + iv - m\Omega_{zz}}}
\end{equation}
Using similar argument like one for $zz$-CNT, the current density for an $ac$-CNT with and without the injection of 
hot electrons are expressed respectively as:
\begin{eqnarray}
j_{zHE}^{ac} = i\frac{4e^2 \gamma_0}{\sqrt{3}n\hbar^2} \sum_{l=1}{r} \sum_{s=1}{f_{rs}\varepsilon_{rs}}
\sum_{m,l=-\infty}{\frac{i^l mlj_m(\beta)j_{m-l}(\beta)I_{m-l}(\beta)\Omega_{ac}}{\omega + iv - m\Omega_{ac}}}\times\nonumber\\
\eta_{ac}\frac{n_0}{2\pi}\sum_{r}\frac{\Omega_{ac}exp(ir\phi)}{(ir\Omega_{zz} + v +\eta\Omega_{ac})}
( exp(-ir\phi^{\prime} - \frac{v}{(v + ir\Omega_{ac})}) + \frac{v}{(v + ir\Omega_{ac})})
\end{eqnarray}
\begin{equation}
j_{z}^{ac} = i\frac{4e^2 \gamma_0}{\sqrt{3}n\hbar^2} \sum_{l=1}{r} \sum_{s=1}{f_{rs}\varepsilon_{rs}}
\sum_{m,l=-\infty}{\frac{i^l mlj_m(\beta)j_{m-l}(\beta)I_{m-l}(\beta)\Omega_{zz}}{\omega + iv - m\Omega_{zz}}}
\end{equation}

\section*{Results and discussion}
We now present a semiclassical theory of electron transport in a CNT under conditions where, in addition 
to the dc field causing a Negative Differential Conductivity (NDC), a similarly strong ac field is present,. Here the ac field plays a two 
fold role: It suppresses the space-charge instability in CNT and simultaneously pumps an energy for 
generation and amplification of THz radiation at higher frequency~\cite{33}. Figure $1$ displays the 
behaviour of the normalized current density ( $J_z = \frac{J_{zHE}^{zz}}{j_0}$, 
where $j_0 = \frac{4e^2 \gamma_0}{\sqrt{3}n\hbar^2}$ ) as a function of the frequency ($\omega$)  of 
$ac$ field for the CNTs stimulated axially with the hot electrons, represented by the nonequilibrium 
parameter $\eta$.
\begin{figure}[h!]
\begin{centering}
\includegraphics[width = 12cm]{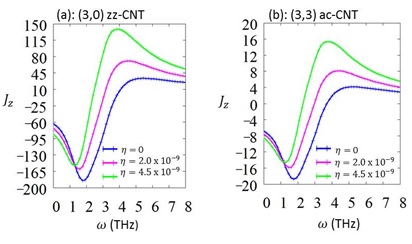} 
 \caption{ A plot of  normalized current density ( $J_z$)  versus frequency of ac field ($\omega$)  as the  
nonequilibrium parameter ($\eta$) increases from $0$ to $4.5 \times 10^{-9}$ for (a) ($3,0$) $zz$-CNT  
and (b) ($3,3$) $ac$-CNT  , $T = 287.5 K$ and $v = 1 THz$} 
\end{centering}
\end{figure}
In the absence of hot electrons ( $\eta = 0$), we  observed  that  the  differential  conductivity  is  
initially negative at zero frequency. With increasing frequency of $ac$ electric field $\omega$ from zero, the 
differential conductivity becomes more negative until a minimum peak is reached at a frequency $\omega$ 
about $1.8 THz$ for both $zz$-CNT and $ac$-CNT .  Then after the differential conductivity turns 
positive when $\omega >1.8 THz$  until the maximum peak is attained at $\omega\approx 4.5 THz$ and then 
decrease when $\omega > 4.5 THz$ for both $zz$-CNT and $ac$-CNT. The Positive Differential Conductivity
(PDC)  is considered as one of the 
conditions for electric stability of the system~\cite{34} and   indicative  for  terahertz  gain 
without  the  small spike or fluctuations of electrons associated with NDC that amplifies,  induces 
space charge accumulation and finally develops into electric field domain~\cite{35}. The electrical 
domain development and transporting induce unstable non uniform electric field distribution, which in 
turn prevents the operation of the Bloch oscillations. 
Thus suppressing domain formation is a prerequisite to observe Bloch oscillations necessary for 
terahertz gain~\cite{35}.
As we increase the nonequilibrium parameter $\eta$ which increases as the rate of hot electrons 
injection increases from $0$ (no hot electrons) to $4.5\times 10^{-9}$ (presence of hot electrons), 
we observed that the minimum peak decreases and shifts to the left (i.e., low frequency). In the 
contrary, the maximum peak increases and also shifts to the left (i.e., low frequency) as shown 
in figure $1$.
In figure $2$, we display the behaviour of normalized current density ($J_z$) as a function of 
frequency of ac field ($\omega$) as the nonequilibrium parameter ($\eta$) is further increased 
to $23.0 \times 10^{-9}$.
\begin{figure}[h!]
\begin{centering}
\includegraphics[width = 12cm]{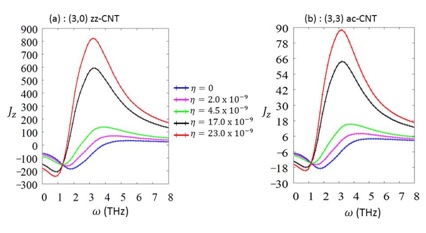} 
 \caption{ A plot of  normalized current density ( $J_z$)  versus frequency of ac field ($\omega$)  as the  
nonequilibrium parameter ($\eta$) increases from  to $23 \times 10^{-9}$ for (a) ($3,0$) $zz$-CNT  
and (b) ($3,3$) $ac$-CNT  , $T = 287.5 K$ and $v = 1 THz$} 
\end{centering}
\end{figure}
As we further increase the nonequilibrium parameter $\eta$  to $23.0 \times 10^{-9}$ ( i.e strong enough 
injection rate), we now noticed that both the  minimum and maximum peaks increase and  shift to the 
left (i.e., low  frequency)  for $\eta \geq 17.0\times 10^{-9}$.  Hence high frequency conductivity of 
strong enough hot electrons in CNTs leads to  increase in both the minimum and maximum peaks of 
normalized current density at lower frequencies as shown in figure $2$. 
To put the above observations in perspective, we display in figures $3$ and $4$, a $3$-dimensional 
behavior of the normalized current density ($J_z$) as a function of the frequency of ac field ($\omega$)  
and nonequilibrium parameter ($\eta$) 
\begin{figure}[h!]
\begin{centering}
\includegraphics[width = 12cm]{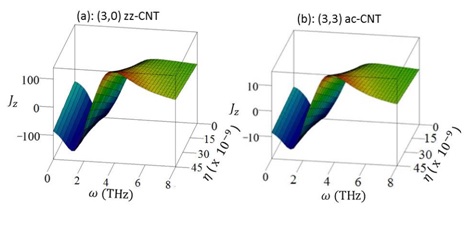} 
 \caption{ A $3D$ plot of  normalized current density ( $J_z$)  versus frequency of ac field ($\omega$)  as the  
nonequilibrium parameter ($\eta$) increases from $0$ to $4.5\times 10^{-9}$ for (a) ($3,0$) $zz$-CNT  
and (b) ($3,3$) $ac$-CNT  , $T = 287.5 K$ and $v = 1 THz$} 
\end{centering}
\end{figure}
\begin{figure}[h!]
\begin{centering}
\includegraphics[width = 12cm]{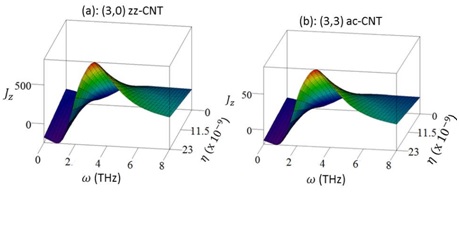} 
 \caption{ A $3D$ plot of  normalized current density ( $J_z$)  versus frequency of ac field ($\omega$)  as the  
nonequilibrium parameter ($\eta$) increases  to $23\times 10^{-9}$ for (a) ($3,0$) $zz$-CNT  
and (b) ($3,3$) $ac$-CNT  , $T = 287.5 K$ and $v = 1 THz$} 
\end{centering}
\end{figure}
In figure $3$, we observed that when nonequilibrium parameter $\eta$ is zero, the minimum peak is the 
greatest at a relative high frequency while the maximum peak is the least also at high frequency. As 
nonequilibrium parameter $\eta$ increases from $0$ to $4.5 \times 10^{-9}$, the  minimum peak gradually 
decreases and shifts towards left ( i.e. low frequency) while the  maximum peak slowly increases and 
also shift towards left ( i.e. low frequency)
In figure $4$,  as nonequilibrium parameter $\eta$ further increases from $0$ to $23.0\times 10^{-9}$, the  
minimum peak initially decreases  and shifts towards left ( i.e. low frequency) and then finally 
increases and shifts towards left ( i.e low frequency) until the highest  minimum peak is attained 
at the lowest frequency.  The trend of the maximum peak as nonequilibrium parameter further increases 
to $23.0 \times 10^{-9}$ remain unchanged.

\section*{Conclusion}
In summary, we have shown theoretically a high frequency conductivity of  hot electrons in a CNT under 
conditions where, in addition to the dc field causing NDC, a similarly strong ac field is applied . 
The applied ac-field plays twofold role of suppressing the space-charge instability in CNT and 
simultaneously pumping an energy for generation and amplification of THz radiation which have enormous 
promising applications in very different areas of science and technology. The generation of this 
radiation occurs at  lower frequency . This is mainly because of increase in both the minimum and 
maximum peaks of normalized current density at lower frequencies as a result of the presence of hot 
electrons .



\end{document}